\documentclass[onecolumn,usenatbib]{mn2e}
\usepackage{times}
\usepackage{epsfig}
\title[Magnetic fields near a rotating black hole]{Dynamical behaviour of the `beads' along the magnetic field lines near a rotating black hole}
\author[X. Cao]{Xinwu Cao\thanks{E-mail:cxw@shao.ac.cn}\\
Shanghai Astronomical Observatory, Chinese Academy of Sciences, 80
Nandan RD, Shanghai£¬200030£¬China}

\begin{document}

\date{Accepted 1997 June 5. February 1. Received 1997 April 4;}

\pagerange{\pageref{firstpage}--\pageref{lastpage}} \pubyear{1997}

\maketitle

\label{firstpage}

\begin{abstract}
The elements of the cold magnetic driven flows behave like beads on
the magnetic field line. The inclination of the field lines at the
surface of the disc plays a crucial role on the nature of the
magnetically driven outflow. For the non-relativistic case, a
centrifugally driven outflow of matter from the disc is possible, if
the poloidal component of the magnetic field makes an angle of less
than a critical 60$^\circ$ with the disc surface. The collimated
flows ejected from active galactic nuclei may probably start from
the region near the black hole. We investigate the dynamical
behavior of the 'beads' on the magnetic field line start from the
disc near a black hole. It is found that the critical angle becomes
larger than 60$^{\circ}$ for the rotating black hole case, which may
imply that the flows are easy accelerated in the inner edge of the
disk surrounding a rotating black hole.

\end{abstract}


\section{Introduction}

For ten years, production of magnetohydrodynamic jets by an
accretion disc has been considered as a promising paradigm to
explain the collimated flows ejected from active galactic nuclei and
more recently by young stellar objects(Blandford and Payne 1982;
Lovelace 1976 and Camezind 1987). In these models the disc is
supposed to be threaded by a vertical magnetic field that extends
through a dense corona, the magnetosphere above it becoming filled
with plasma which can be flung out centrifugally by the magnetic
field lines. In magnetically driven flows, the strength and
distribution of the magnetic flux at the disc surface plays an
important role as a boundary condition for the problem. The angular
momentum loss rate of the flow depends mostly on the strength of the
field. The inclination of the field lines at the surface  strongly
influences the way in which the flow is accelerated. The elements of
relatively tenuous gas on the magnetic field lines behave like beads
on a rigid wire. Blandford and Payne (1982) point out that the beads
can be launched centrifugally from a Keplerian accretion disc if the
poloidal field direction is inclined at an angle less than
60$^\circ$ to the radial direction. Cao and Spruit (1994)
investigate in some detail how the flow properties change with the
field inclination at the disc surface. At low inclination, they find
a magnetically driven circulation along the disc surface rather than
a high density flow. The jets observed in some active galactic
nuclei are believed to be accelerated close to the black hole. In
this paper we analyze the dynamical features of the beads along the
magnetic field line near a rotating black hole in general
relativistic frame. The basic equations for the problem are listed
in Sect. 2. Sect. 3 contains a brief discussion.

\section{Basic equations}

The behaviour of flows flung out centrifugally by the magnetic
fields is like beads on rigid wires. If the 'beads' start from rest
at the disc surface and are carried with constant angular velocity
by the 'wire', then the effective potential for  a bead on a wire,
corotating with the Keplerian angular velocity at a radius $r_{d}$,
is

\begin{equation}
\Psi_{eff}(r, z)=-{{GM}\over{r_{d}}} \left[ {1\over 2} \left(
{r\over{r_{d}}} \right)^{2} +{ {r_{d}}\over {(r^{2}+z^{2})^{1/2}}
}\right],
\end{equation}
where the Newtonian gravitational potential is used in the derivation.
After some mathematical derivations, one finds that the equilibrium at
$r=r_{d}$ is unstable if the projection of the wire on the meridional plane
makes an angle of less than 60$^\circ$ with the equatorial plane (Blandford
and Payne 1982). The critical angle of 60$^\circ$  plays an important role
on the jet formation. This mechanical analogy does exhibit the essential
dynamical feature of centrifugally driven flows not very close to a compact
central object. Here we investigate the dynamical properties of beads on the
magnetic field line near a rotating black hole.

The space-time generated by a rotating black hole can be represented by
Kerr metric in spherical coordinates

$$ds^{2}=-{{\triangle}\over {\rho^{2}}}[dt-a\sin^{2}\theta d\phi]^{2}
+{{\sin^{2}\theta}\over {\rho^2}}[(r^{2}+a^{2})d\phi-adt]^{2}
+{{\rho^2}\over{\triangle}}dr^{2}+\rho^{2}d\theta^{2}, \eqno(2)$$
where
$$\triangle\equiv r^{2}-2Mr+a^{2}, $$
$$\rho^{2}\equiv r^{2}+a^{2}\cos^{2}\theta,$$
$a$ is the angular momentum per unit mass of the black hole, $M$ is
the mass of the black hole (setting $G=c=1$).

The radial motion of a free particle close to a Kerr black hole is governed
by the equation:
$${ {d^{2}r}\over {d\tau^{2}} }
+{1\over 2}g^{rr}g_{rr}^{\prime}\left( { {dr}\over{d\tau} }
\right)^{2} -{1\over 2}g^{rr}g_{\theta\theta}^{\prime}\left(
{{d\theta}\over{d\tau}} \right)^{2}-{1\over
2}g^{rr}g_{\phi\phi}^{\prime} \left ({ {d\phi}\over {d\tau}
}\right)^{2} -{1\over 2}g^{rr}g_{tt}^{\prime}\left({{dt}\over
{d\tau}} \right)^{2} -g^{rr}g_{t\phi}^{\prime} { {d\phi} \over
{d\tau} } { {dt} \over {d\tau}} =0,  \eqno(3)$$ where the metric
coefficients $g^{rr}$, $g_{\mu\nu}$, are given by Eq. (2), and
${g_{\mu\nu}^{\prime}} = {{\partial g_{\mu\nu}}\over {\partial r}}$.

Since we only intend to investigate the instability condition of the
particle rotating with the Keplerian angular velocity near the disc
surface, the term $-{1\over 2}g^{rr}g_{\theta\theta}^{\prime}
\left({{d\theta}\over{d\tau}} \right)^{2}$ in Eq. (3) could be
omitted, similar to the non-relativistic approach. Combining Eqs.
(2) and (3), we obtain the equation describing the radial motion of
the beads:
$${ { {du^{r}}^{2} } \over {dr} }+{ {g_{rr}^{\prime}} \over {g_{rr}} }
{u^{r}}^{2}
+{{g_{\phi\phi}^{\prime}\Omega^{2}+2g_{t\phi}^{\prime}\Omega+g_{tt}^{\prime}}
\over {g_{\phi\phi}\Omega^{2}+2g_{t\phi}\Omega+g_{tt}}}{u^{r}}^{2}
-{g^{rr}({g_{\phi\phi}^{\prime}\Omega^{2}+2g_{t\phi}^{\prime}\Omega+g_{tt}
^{\prime}}) \over
{g_{\phi\phi}\Omega^{2}+2g_{t\phi}\Omega+g_{tt}}}=0, \eqno(4)$$
where ${u^{r}}^{2}={{dr}\over {d\tau}}$, $\Omega={{d\phi}\over
{dt}}$.

The beads move along a rigid magnetic field line have the constant
angular velocity measured by a distant observer
$\Omega={{d\phi}\over {dt}}$. For the Keplerian accretion disc,
$\Omega=\Omega_{K}$, given by
$$\Omega_{K}={{u^{\phi}}\over {u^{t}}}, \eqno(5)$$
where $u^{\phi}$, $u^t$ are velocities of a free particle moving in
a circular Keplerian orbit. Using the four-velocity presented in
Bardeen et al. (1972), we finally obtain Keplerian angular velocity
measured by a distant observer
$$\Omega_{K}=\pm {{M^{1/2}}\over {r_{d}^{3/2}\pm aM^{1/2}}}, \eqno(6)$$
where the upper sign is for the direct orbit while  the lower sign
the retrograde orbit, $r_{d}$ is the radial position of the
footpoint of the field line. The radius of minimum stable circular
orbit $r_{ms}$ surrounding a rotating black is
$$r_{ms}=M\{3+Z_{2}\mp [(3-Z_{1})(3+Z_{1}+2Z_{2}) ]^{1/2}\}, \eqno(7)$$
where
$$Z_{1}\equiv 1+\left(1-{{a^2}\over {M^2}}\right)^{1/3} \left [(1+{a\over M}
)^{1/3}+(1-{a\over M})^{1/3}\right ], $$
$$Z_{2}\equiv \left( {{3a^2}\over {M^2}}+Z_{1}^{2}\right )^{1/2},$$
the upper sign for the direct orbit.

Integrating Eq. (4), we can get the effective potential for a bead
on a magnetic field line corotating with the Keplerian angular
velocity $\Omega_{K}$:
$$\Psi_{eff}=-{1\over {g_{rr}}}-{C \over {g_{rr}({g_{\phi\phi}}\Omega_{K}^{2}
+2{g_{t\phi}}\Omega_{K}+{g_{tt}}) }}, \eqno(8)$$ where the integral
constant $C$ is determined by the boundary condition
$$\Psi_{eff}=0, ~~~~~~at~~~~ r=r_{d},~~~ \theta={\pi\over 2}. \eqno(9)$$
Now, the instability condition could then be derived from the second
derivative of the effective potential. After some mathematical
deductions, we get the critical angle $\alpha_{crit}$ of the
magnetic line at the surface of the disc as follows
$$\alpha_{crit}=\arctan
\left \{{ { r_{d}^{2} [ (2
{r_d^{3}}+4M{a^{2}}){\Omega_{K}^2}-8Ma\Omega_{K} +4M ]} \over
{(2{r_{d}^{5}}+2a^{2}{r_{d}^{3}}+8Ma^{2}r_{d}^{2}+4Ma^{4})\Omega_{K}^2-
8Ma({r_{d}^{2}}+a^{2})\Omega_{K}+4Ma^2}} \right \} ^{1/2}.
\eqno(10)$$ In the non-rotating black hole case, $a=0$, we have
$\alpha_{crit}= 60^{\circ}$, which is same as that obtained by
Blandford and Payne (1982) for the non-relativistic case. In Fig. 1
we depict the relations between the critical angle and radial
position of the footpoint of the magnetic line for the rotating
black hole with different angular momentum $a$.

\begin{figure}
\centerline{\psfig{figure=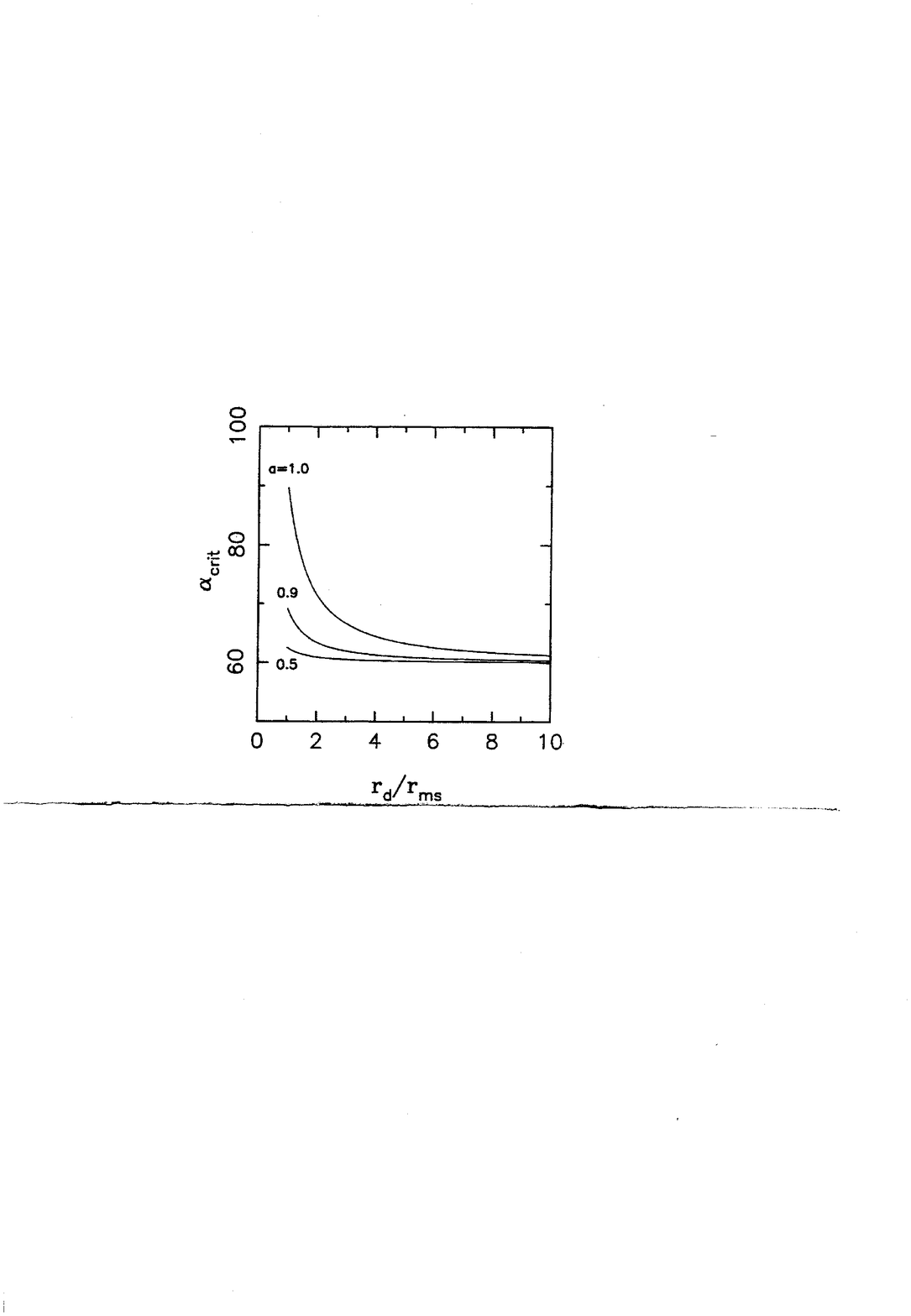,width=12.5cm,height=15.5cm}}
\caption{The critical angle $\theta_{crit}$ vs. radial position of
the magnetic field footpoint at the disc surface with respect to the
different rotating black hole angular momentum, $a=0.5, 0.9, 1.0
M$.}
\end{figure}

\section{Discussion}

The configuration and strength of the magnetic field of the disc plays a
crucial role on the dynamical properties of the centrifugally driven
magnetohydrodynamic flow. Assuming that the field configuration above the
disc is close to a potential field, the shape of the disc field is determined
entirely by the magnetic field $B_{z}$ at the disc surface. The elements of
the cold magnetic driven flows behave like beads on the magnetic field
lines. Whether the matter at the disc surface can be flung out mainly depends
on the inclination angle of the field line at the disc surface (Blandford and
Payne 1982). Blandford and Payne (1982) point out that the flow can be
centrifugally driven if the field lines inclined less than the critical
angle of 60$^\circ$ with respect to the radial direction. The space near
the inner region of the disc is a 'dead  zone', where the magnetic field line
is too steep, the flow not being able to overcome a barrier of gravitational
potential.

We investigate the dynamical properties of the bead on the
magnetic fields threading the rotating disc around a Kerr black
hole.  We have the same critical angle 60$^\circ$ for the non-rotating
black hole ($a=0$) as Newtonian case.
Even if in the region very close to a Schwarzschild black
hole, the flow could not be centrifugally flung out if the field lines
inclined greater than the critical angle 60$^\circ$ with respect to the
radial direction. Abramowicz(1990) point out that the centrifugal force
reverses sign at $r=3M$. For the free particle in circular motion, the
minimum radius is $4M$ (unstable). The results obtained here could not be
extended to the region $r=3M$. Cao(1995) study the same problem by using
a pseudo-Newtonian potential to simulate general relativistic effects.
They find that the critical angle becomes larger in the region close to
a non-rotating black hole. The reason is that the angular velocity derived
from the pseudo-Newtonian potential simulates ${{d\phi}\over {d\tau}}$,
not ${{d\phi}\over {dt}}$. In general relativistic frame, ${{d\phi}
\over {d\tau}}$ and ${{d\phi}\over {dt}}$ have different values. The
beads move along a rigid  magnetic field line have a constant angular
velocity ${{d\phi}\over {dt}}$ measured by a distant observer.

The results are shown in Fig. 1 for the rotating black holes. The critical
angle $\theta_{crit}$ increases with the reduction of the radius of the
magnetic field line footpoint. For the beads at the minimum stable radius
of an extreme Kerr black hole, the critical angle $\theta_{crit}$ could be as
large as $90^\circ$, which may imply that the flows could be centrifugally
accelerated to high velocities even by the magnetic field lines with low
inclination angles very close to the black hole. We note that the critical
angle falls rapidly in the cases of relative slowly rotating black hole,
for example, $\theta_{crit} \sim 70^{\circ}$ at the inner edge of the disc
for $a=0.9$. The results obtained here strongly imply that the rotating
black hole will be helpful on the centrifugally acceleration of flows
very close to the black hole. The 'dead zone' for accelerating over the
the inner region of the disc will disappear for the rapidly rotating black
hole with an appropriate disc magnetic field configuration. The further
investigation on this problem using the full relativistic MHD approach
is necessary.

We also study the case that the matter in the disc is in the retrograde orbit
around the black hole, though it is not sure whether such disc exists in
reality. The results show that the critical angle is slightly influenced
by the rotating black hole, the critical angle is always less than, but very
close to 60$^\circ$. The critical angle at the minimum stable radius
for the extreme Kerr black hole is about 57$^\circ$, almost same as the
non-rotating case.


\end{document}